\documentclass[11pt,a4paper]{article}
\usepackage[T1]{fontenc}
\usepackage[utf8]{inputenc}
\usepackage{mathptmx}
\usepackage{microtype}
\usepackage{amsmath,amssymb,amsthm}
\usepackage{mathtools}
\usepackage{bm}
\usepackage{geometry}
\usepackage{graphicx}
\graphicspath{{weak_shock_figures/}}
\usepackage{xcolor}
\usepackage{booktabs}
\usepackage{float}
\usepackage[authoryear]{natbib}
\usepackage{hyperref}
\hypersetup{hidelinks}
\usepackage{cleveref}
\usepackage{setspace}
\usepackage{fancyhdr}
\usepackage{titlesec}
\usepackage{abstract}
\theoremstyle{plain}
\newtheorem{theorem}{Theorem}[section]
\newtheorem{proposition}[theorem]{Proposition}

\theoremstyle{definition}

\newtheorem{remark}[theorem]{Remark}
\newcommand{\al}{\alpha}
\newcommand{\lam}{\lambda}

\newcommand{\dd}{\mathrm{d}}

\newcommand{\Mach}{M}
\newcommand{\OO}{\mathcal{O}}
\newcommand{\TP}{\mathcal{T}}
\newcommand{\utsd}{\textsc{utsd}}
\newcommand{\half}{\tfrac12}
\long\def\revone#1{#1}
\long\def\revtwo#1{#1}

\titleformat{\section}{\large\bfseries}{\thesection.}{0.6em}{}
\titleformat{\subsection}{\normalsize\bfseries}{\thesubsection.}{0.6em}{}
\titleformat{\subsubsection}{\normalsize\itshape}{\thesubsubsection.}{0.6em}{}

\begin{document}
% ------- Title block -------------------------------------------
\begin{center}
  {\LARGE\bfseries Distinguished Scaling and UTSD Structure\\[4pt]
    in Weak Shock Reflection at Nearly Glancing Incidence}\\[1.4em]
  {\large Justin Kin Jun Hew$^{1}$}\\[0.5em]
  {\small $^{1}$ACCESS-NRI (Australia's Climate Simulator), Australian National University,
    Canberra, ACT 2601, Australia}\\[0.4em]
  {\small \texttt{justinkinjun.hew@anu.edu.au}}\\[1.4em]
  \rule{0.55\linewidth}{0.5pt}
\end{center}

\begin{abstract}
\noindent
\revone{We study weak shock reflection from a rigid wall in the joint limit of weak
shock strength and nearly glancing incidence.  In the distinguished scaling
$\Mach=1+\lambda\alpha^2$, the inner reflection region is governed by the
unsteady transonic small-disturbance (\utsd) equation and is controlled, to
leading order, by the single parameter $a_0=1/(2\sqrt{\lambda})$, independent
of the ratio of specific heats $\gamma$. Thus the known \utsd{} detachment
value $a_d=\sqrt2$ corresponds in this scaling to $\lambda_d=1/8$, with
Guderley--Mach reflection for $\lambda>1/8$. The physical trajectory angle is
obtained by multiplying the canonical \utsd{} trajectory function $g(a)$ by
the Mach-number strength scale $\delta=\sqrt{2(\Mach^2-1)}$, so that
$\chi_{\rm phys}=\delta g(a)+O(\delta^2)=2\sqrt{\lambda}\,\alpha g(a_0)+O(\alpha^3)$.
We rederive the self-similar \utsd{} reduction, sonic parabola, and shock
polar in order to make the convention and the detachment map self-contained.
We also record a formal adjoint solvability expression for the first correction
$H(a;\gamma)$, while specifying the free-boundary data required to evaluate it.
Finally, a time-marching solver for the full leading-order canonical \utsd{}
system is benchmarked at $a_0=0.5$: retaining the transverse compression
$u>1$ gives a $u=0.5$ contour location consistent with the Hunter--Tesdall
triple-point benchmark. This computation is used only as a leading-order
benchmark, not as a substitute for an adaptive self-similar Guderley
free-boundary solver.}
\end{abstract}
\bigskip

% ===================================================================
\section{Introduction}
\label{sec:intro}
% ===================================================================
The reflection of a planar shock wave from a rigid plane wall is a
foundational problem of compressible-flow theory
\citep{VonNeumann1943,Courant1948,BenDor2007}. For steep incidence the flow
admits a two-shock \emph{regular reflection} (RR); below a critical angle it
transitions to \emph{Mach reflection} (MR), with a Mach stem bridging the
incident and reflected shocks at a triple point $\TP$ above the wall
\citep{Henderson1987,BenDor2007}.

For weak shocks the transition occurs in a thin-wedge/weak-shock regime in
which the classical RR/MR criteria fail---the \emph{von~Neumann paradox}
\citep{VonNeumann1943,Colella1990}. Experiments \citep{Skews2005} and
high-resolution computations \citep{TesdallHunter2002,VasilevKraiko1999} show
that the paradox is resolved, following Guderley's conjecture
\citep{Guderley1947}, by replacing the sharp triple point with a self-similar
sequence of supersonic patches---a \emph{Guderley--Mach reflection}. The local
flow is governed not by the Rankine--Hugoniot system but by the unsteady
transonic small-disturbance (\utsd) equation
\citep{HunterBrio2000,HunterTesdall2004,Brio2001}, which in canonical form
contains no $\gamma$.

\revone{This paper is a technical note on the scaling connection between the outer
shock-polar description and the inner \utsd{} reflection problem. It does not
claim a new Guderley reflection mechanism or a complete matched-asymptotic
solution. The canonical \utsd{} equation, the mixed-type self-similar Guderley
structure, the detachment value $a_d=\sqrt2$, and the numerical nature of the
trajectory function $g(a)$ are known from the work of Hunter, Brio, Tesdall,
and collaborators. Here these results are used as the inner canonical problem
and, where useful, rederived to fix conventions. The new contribution is the
explicit translation of that canonical problem into the distinguished physical
scaling $\Mach=1+\lambda\alpha^2$, including the resulting $\gamma$-free inner
parameter $a_0=1/(2\sqrt\lambda)$, the corresponding threshold $\lambda_d=1/8$,
and the precise angular normalisation by the Mach-number strength scale.}

\revone{For clarity, the role of each part of the paper is as follows:}
\begin{enumerate}
  \item \revone{The known inner problem is the Hunter--Brio/Hunter--Tesdall \utsd{}
        reflection problem; we use it as the canonical leading-order model.}
  \item \revone{We rederive the self-similar reduction, sonic parabola, and \utsd{}
        shock polar to make the angle convention and the detachment threshold
        self-contained.}
  \item We also give the ordinary outer flow-deflection expansion and an
        auxiliary parabolic-fan calculation showing why an elementary
        higher-order closure cannot be extracted from that local ansatz alone.
  \item \revone{The main scaling result is that, for $\Mach=1+\lambda\alpha^2$,
        the limiting inner parameter is $a_0=1/(2\sqrt\lambda)$ and no
        independent $\gamma$-dependence enters the leading canonical \utsd{}
        problem.}
  \item \revone{The physical trajectory angle is the canonical \utsd{} trajectory
        multiplied by $\delta=\sqrt{2(\Mach^2-1)}$, so
        $\chi_{\rm phys}=2\sqrt\lambda\,\alpha\,g(a_0)+O(\alpha^3)$.}
  \item \revone{The first correction is not computed as a curve. Instead, we define
        the coefficient $H(a;\gamma)$ through the formal adjoint solvability
        condition for the full linearised Guderley free-boundary problem.}
  \item \revtwo{The pilot numerical calculation is a leading-order time-marching
        benchmark for the canonical \utsd{} equations, not an adaptive
        self-similar free-boundary calculation. Its limitations are made
        explicit in \Cref{sec:pilot}.}
\end{enumerate}

% ===================================================================
\section{Outer Expansion and the Control Parameter}
\label{sec:outer}
% ===================================================================
\subsection{Geometry, scaling, and convention}
\label{sec:convention}
\revone{We work in the self-similar frame attached to the reflection point, in which
all quantities depend on $(X,Y)=(x/t,y/t)$. Throughout the paper $\alpha$
denotes the small glancing angle between the incident shock and the wall; this
is the same small angle that is sometimes described as the thin-wedge angle in
the equivalent wedge formulation. We avoid using a separate incidence-angle
notation. Let $\Mach$ be the incident-shock Mach number. The weak-shock,
nearly-glancing regime is the transitional limit \citep{HunterTesdall2004}}
\begin{equation}
  \Mach\to1,\quad \al\to0,\qquad \lam\equiv\frac{\Mach-1}{\al^2}\ \text{fixed},
  \label{eq:scaling}
\end{equation}
so that the shock-strength parameter has the explicit expansion
\begin{align}
  \beta &:= \Mach^2-1
  =(1+\lam\al^2)^2-1  \notag\\
  &=2\lam\al^2+\lam^2\al^4
   =2\lam\al^2\left(1+\frac{\lam}{2}\al^2\right).
  \label{eq:strength-explicit}
\end{align}
Thus $\lambda$ measures shock strength relative to the square of the glancing
angle. This single convention is used throughout. In particular, all formulae
below are to be read in the distinguished limit $\al\to0$ with $\lambda$ fixed;
finite-$\alpha$ equalities inherit the displayed $\OO(\al^2)$ corrections.

It is also useful to state explicitly which Mach number enters the outer
Rankine--Hugoniot relations. In the pseudo-steady frame attached to the
reflection point, the incident shock may be viewed as an oblique shock with
upstream Mach number $\widetilde M=\Mach/\sin\alpha$ and wave angle
$\beta=\alpha$. Its normal Mach number is therefore
$\widetilde M\sin\beta=\Mach$. Hence the pressure jump and flow deflection
are governed by the same shock Mach number $\Mach$ used in \eqref{eq:scaling};
no second Mach-number convention is introduced.

\subsection{Incident-shock strength and pressure jump}
The incident shock is a shock of strength
$\beta=\Mach^2-1=2\lam\al^2+\OO(\al^4)$, a purely kinematic quantity independent of
$\gamma$.  The usual weak normal-shock pressure relation gives
\begin{equation}
  \frac{p_1-p_0}{p_0}
  =\frac{2\gamma}{\gamma+1}\bigl(\Mach^2-1\bigr)+\OO(\beta^2),
\end{equation}
so that, in the nondimensional pressure variable used by the small-disturbance
scaling,
\begin{equation}
  \frac{p_1-p_0}{\gamma p_0}=\frac{2}{\gamma+1}\bigl(\Mach^2-1\bigr)
  =\frac{4\lam}{\gamma+1}\,\al^2
    \left(1+\frac{\lambda}{2}\al^2+\OO(\al^4)\right).
  \label{eq:pjump}
\end{equation}
By contrast, the canonical UTSD strength is obtained by removing the glancing
scale,
\begin{equation}
  \frac{2(\Mach^2-1)}{\al^2}=4\lam+2\lam^2\al^2,
  \label{eq:canonical-strength}
\end{equation}
which is independent of $\gamma$. This elementary distinction is the source
of the apparent disappearance of $\gamma$ from the inner problem: $\gamma$ is
present in the physical pressure jump \eqref{eq:pjump}, but not in the
canonical UTSD parameterisation \eqref{eq:canonical-strength}.

\subsection{Outer flow deflection}
\label{sec:outer-deflection}
The same convention gives a useful check on the outer shock polar. Let
$\vartheta$ denote the physical deflection of the post-incident-shock flow in
the pseudo-steady frame. With upstream Mach number
$\widetilde M=\Mach/\sin\alpha$ and wave angle $\beta=\alpha$, the standard
oblique-shock relation gives
\begin{equation}
  \tan\vartheta=
  \frac{2\cot\beta\, (\widetilde M^2\sin^2\beta-1)}
       {\widetilde M^2(\gamma+\cos 2\beta)+2},
  \qquad \widetilde M^2\sin^2\beta-1=\Mach^2-1 .
  \label{eq:outer-defl-tan}
\end{equation}
Expanding \eqref{eq:outer-defl-tan} with $\Mach=1+\lambda\alpha^2$ gives
\begin{equation}
  \vartheta=
  \frac{4\lambda}{\gamma+1}\,\alpha^3
  \left[1-\left(\frac{3\lambda}{2}+\frac23\right)\alpha^2
  +\OO(\alpha^4)\right].
  \label{eq:outer-defl}
\end{equation}
Thus the ordinary post-shock flow deflection is $\OO(\alpha^3)$ in the
present nearly-glancing weak-shock scaling. This should not be confused with
the small physical trajectory angle $\chi_{\rm phys}$ of the triple point,
whose inner angular scale is $\delta=\sqrt{2(\Mach^2-1)}=\OO(\alpha)$ and
is discussed in \Cref{sec:angle-scaling}. Equation \eqref{eq:outer-defl} is
included to make the outer shock-polar convention explicit and to show that the
weak oblique-shock polar is consistent with the distinguished balance.

\subsection{The single, $\gamma$-free control parameter}
\label{sec:param}
\begin{figure}[t]
  \centering
  \includegraphics[width=0.74\linewidth]{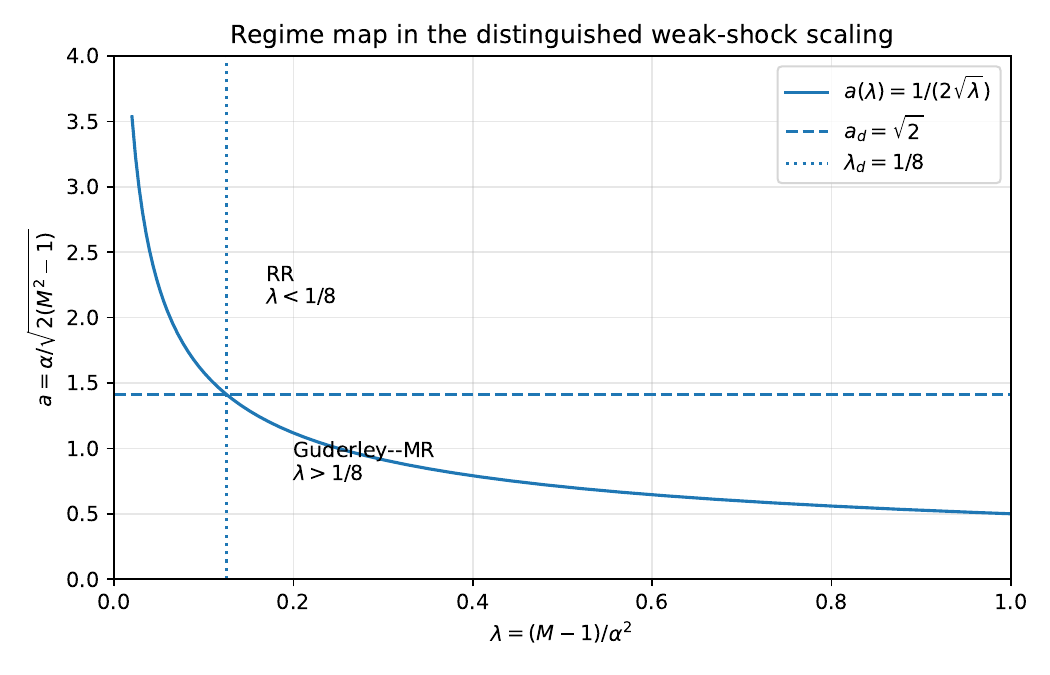}
  \caption{\revone{Regime map illustrating the chosen distinguished weak-shock scaling convention $\Mach-1=\lambda\alpha^2$.}
  The limiting Hunter--Tesdall control parameter is $a_0=1/(2\sqrt{\lambda})$. The detachment value
  $a_d=\sqrt2$ maps to $\lambda_d=1/8$, separating regular reflection
  $(\lambda<1/8)$ from the Guderley--Mach reflection regime $(\lambda>1/8)$.}
  \label{fig:regime-map}
\end{figure}

\begin{figure}[t]
  \centering
  \includegraphics[width=0.74\linewidth]{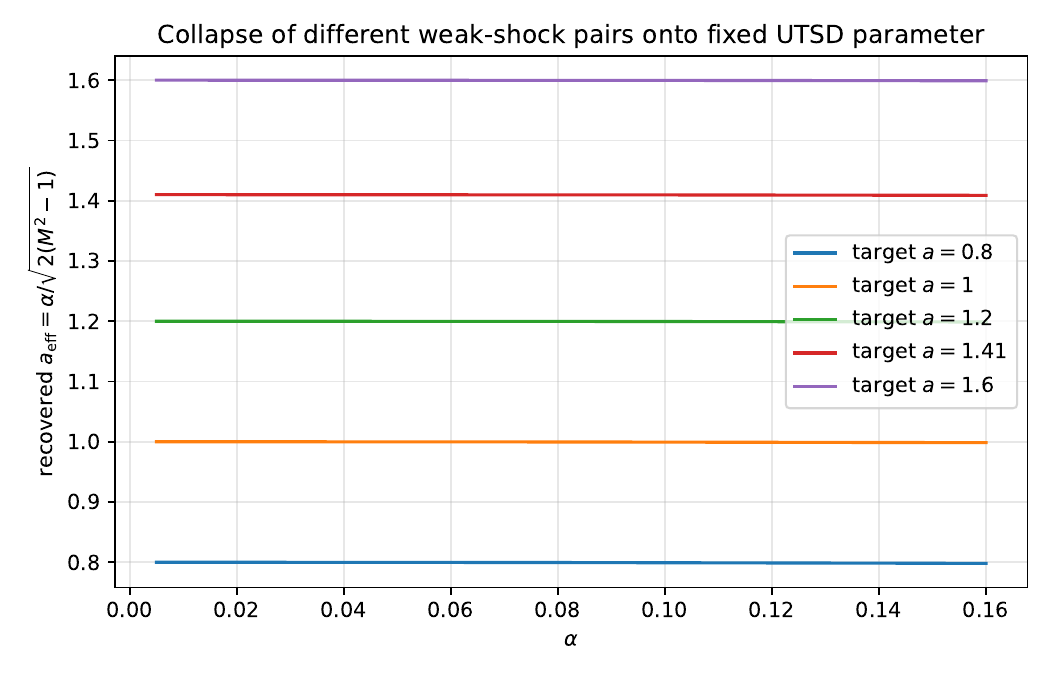}
  \caption{\revone{Scaling-collapse diagnostic illustrating the chosen $a$--$\lambda$ convention.} For several target values of the UTSD control
  parameter $a$, the curves show the recovered value
  $a_{\mathrm{eff}}=\alpha/\sqrt{2(\Mach^2-1)}$ from weak-shock pairs satisfying
  $\Mach-1=\lambda\alpha^2$ with $\lambda=1/(4a^2)$. The small residual variation is the
  expected $\mathcal{O}(\alpha^2)$ effect from using the exact $\Mach^2-1$ rather than only
  its leading approximation $2(\Mach-1)$.}
  \label{fig:parameter-collapse}
\end{figure}

The transitional-limit matched asymptotics \citep{HunterBrio2000,HunterTesdall2004}
reduce the reflection to the \utsd{} shock-reflection problem
(\S\ref{sec:inner}), controlled by the single parameter
\begin{align}
  a&=\frac{\al}{\sqrt{2(\Mach^2-1)}}
   =\frac{\al}{\sqrt{4\lam\al^2+2\lam^2\al^4}} \notag\\
   &=\frac{1}{2\sqrt{\lam}}\left(1+\frac{\lam}{2}\al^2\right)^{-1/2}
    =\frac{1}{2\sqrt{\lam}}\left(1-\frac{\lam}{4}\al^2+
      \OO(\al^4)\right). \label{eq:a-expansion}
\end{align}
Thus
\begin{equation}
  \boxed{\ a_0:=\lim_{\al\to0}a=\frac{1}{2\sqrt{\lam}}\ }.
  \label{eq:a}
\end{equation}
The limiting value $a_0$ is held fixed in \eqref{eq:scaling}. The geometry (numerator $\al$) and
the strength (denominator $\sqrt{2(\Mach^2-1)}$) enter separately; combining
them, the limiting parameter $a_0$ is independent of both $\gamma$ and $\al$.
The map between $\lambda$ and the Hunter--Tesdall control parameter is shown
in \Cref{fig:regime-map}, and the collapse of weak-shock pairs onto fixed $a$ is
shown in \Cref{fig:parameter-collapse}.
\begin{proposition}[Single $\gamma$-free inner parameter]
\label{prop:single}
The inner self-similar \utsd{} reflection problem depends on $(\gamma,\lam)$
only through $\lam$, equivalently through $a_0=1/(2\sqrt\lam)$; it does not
depend on $\gamma$. If $g(a)$ denotes the canonical UTSD trajectory function,
then the physical small angle is obtained by multiplying by the Mach-number
strength scale $\delta=\sqrt{2(\Mach^2-1)}$. Thus $\gamma$ does not enter as
an independent inner parameter or as a separate prefactor in this convention.
\end{proposition}

\subsection{Physical angular normalisation}
\label{sec:angle-scaling}
\revone{The canonical UTSD trajectory function is a scaled angular quantity, so the
overlap between the physical outer variables and the inner \utsd{} variables
must specify which angle is being measured. In what follows $\chi_{\rm phys}$
denotes the small physical angle made in the physical self-similar plane by the
leading triple-point/Mach-stem trajectory relative to the wall-attached
reflection point. Define}
\begin{equation}
  \delta=\sqrt{2(\Mach^2-1)} .
  \label{eq:delta}
\end{equation}
\revone{The Hunter--Tesdall parameter is $a=\alpha/\delta$. Thus, in the overlap
region, physical angular variations of order $\alpha$ are represented inside
the canonical \utsd{} problem as order-one angular variables scaled by
$\delta$. This can be seen directly from slopes in the self-similar plane. A
physical ray making a small angle $\theta_{\rm phys}$ with the wall has
$Y/X=\tan\theta_{\rm phys}=\theta_{\rm phys}+\OO(\theta_{\rm phys}^3)$. In the
inner UTSD scaling the corresponding canonical angular variable is
$\theta_{\rm phys}/\delta$; in particular, the incident geometry maps to
$a=\alpha/\delta$. Therefore an order-one canonical trajectory slope or angle
$g(a)$ corresponds in physical variables to the small angle}
\begin{equation}
  \chi_{\rm phys}=\delta\,g(a)+\OO(\delta^2).
  \label{eq:chi-delta}
\end{equation}
\revone{The error term represents next-order corrections from the Euler-to-\utsd{}
expansion and from the outer shock-polar matching, not a second independent
inner parameter. This is the overlap relation used below.}
Under the distinguished scaling $\Mach=1+\lambda\alpha^2$,
\begin{align}
  \delta
  &=\sqrt{2(\Mach^2-1)}
    =\sqrt{4\lambda\alpha^2+2\lambda^2\alpha^4} \notag\\
  &=2\sqrt{\lambda}\,\alpha
    \left(1+\frac{\lambda}{4}\alpha^2+\OO(\alpha^4)\right),
  \label{eq:delta-expansion}
\end{align}
and therefore
\begin{equation}
  \boxed{\chi_{\rm phys}=2\sqrt{\lambda}\,\alpha\,g(a_0)+\OO(\alpha^3),
  \qquad a_0=\frac{1}{2\sqrt{\lambda}}.\ }
  \label{eq:chi-phys}
\end{equation}
Equivalently,
\begin{equation}
  \frac{\chi_{\rm phys}}{\alpha}=\frac{g(a_0)}{a_0}+\OO(\alpha^2).
  \label{eq:chi-over-alpha}
\end{equation}
No additional $\gamma$-dependent prefactor appears in this Mach-number-based
scaling. A $\gamma$-dependent factor arises only if the shock strength is
parameterised by the physical pressure jump rather than by $\Mach^2-1$; using
\eqref{eq:pjump},
\begin{equation}
  \sqrt{2(\Mach^2-1)}=
  \sqrt{(\gamma+1)\,\frac{p_1-p_0}{\gamma p_0}},
  \label{eq:pressure-normalisation}
\end{equation}
so an apparent factor $\sqrt{\gamma+1}$ is a change of normalisation, not a
dependence of the canonical UTSD inner problem.

\begin{remark}
A tempting normal-Mach-number parametrisation would use the apparent strength
$\Mach_n^2-1=(2\lam-1)\al^2$ and would suggest a lower bound
$\lam>\half$. That choice folds the obliquity factor $\cos^2\al$ into the
strength, double-counting the geometry that the control parameter \eqref{eq:a}
keeps separate; it is therefore inconsistent with the parametrisation of
\citet{HunterTesdall2004}.
\end{remark}

% ===================================================================
\section{Inner Expansion: The \utsd{} Reflection Problem}
\label{sec:inner}
% ===================================================================
\subsection{The canonical equation and the reflection problem}
The transitional limit yields the canonical \utsd{} equation, which contains
no $\gamma$,
\begin{equation}
  u_t+\Bigl(\tfrac12u^2\Bigr)_x+v_y=0,\qquad u_y=v_x,
  \label{eq:canonUTSD}
\end{equation}
together with the self-similar shock-reflection initial--boundary value problem
\citep{HunterTesdall2004}
\begin{equation}
  u(x,y,0)=
  \begin{cases}1,&x<ay,\\ 0,&x>ay,\end{cases}
  \qquad v(x,0,t)=0,\qquad v\to0\ \text{as}\ x\to+\infty,
  \label{eq:IBVP}
\end{equation}
with $a$ the parameter \eqref{eq:a}.

\subsection{The self-similar reduction and the sonic line}
Under $u=u(\xi,\eta)$, $v=v(\xi,\eta)$ with $\xi=x/t$, $\eta=y/t$, the chain rule gives
\begin{equation}
  u_t=-\frac{\xi}{t}u_\xi-\frac{\eta}{t}u_\eta,
  \qquad
  \left(\frac12u^2\right)_x=\frac{1}{t}u u_\xi,
  \qquad
  v_y=\frac{1}{t}v_\eta.
\end{equation}
Multiplying the first UTSD equation by $t$ gives
\begin{equation}
  (u-\xi)\,u_\xi-\eta\,u_\eta+v_\eta=0.
\end{equation}
Similarly, $u_y=t^{-1}u_\eta$ and $v_x=t^{-1}v_\xi$, so the compatibility
condition becomes
\begin{equation}
  (u-\xi)\,u_\xi-\eta\,u_\eta+v_\eta=0,\qquad u_\eta=v_\xi.
  \label{eq:ssUTSD}
\end{equation}
The type-change curve follows by introducing a local potential
$\Phi(\xi,\eta)$ with $u=\Phi_\xi$ and $v=\Phi_\eta$, possible away from
shocks because $u_\eta=v_\xi$.  The principal part of the resulting second-order
self-similar equation is
\begin{equation}
  (u-\xi)\Phi_{\xi\xi}-\eta\Phi_{\xi\eta}+\Phi_{\eta\eta}.
\end{equation}
For a second-order equation $A\Phi_{\xi\xi}+2B\Phi_{\xi\eta}+C\Phi_{\eta\eta}$,
the discriminant is $B^2-AC$.  Here
\begin{equation}
  B^2-AC=\frac{\eta^2}{4}-(u-\xi),
\end{equation}
so the sonic parabola is
\begin{equation}
  u=\xi+\frac14\eta^2,
  \label{eq:sonic-parabola}
\end{equation}
with \eqref{eq:ssUTSD} hyperbolic for $u-\xi<\eta^2/4$ and elliptic otherwise.
The global self-similar solution of \eqref{eq:ssUTSD} is not a single smooth
fan.  A simple parabolic local fan can nevertheless be useful as a matching
calculation and as a test of possible higher-order closures.  We record that
calculation in \Cref{app:parabolic-fan}; it satisfies the compatibility
constraint exactly but does not replace the mixed-type Guderley solution.

\subsection{The leading-order field is the Guderley patch structure}
Following \citet{Guderley1947} and the numerical solutions of
\citet{HunterBrio2000,TesdallHunter2002}---confirmed by
\citet{TesdallSandersKeyfitz2006} and \citet{VasilevKraiko1999} and observed
experimentally by \citet{Skews2005}---the self-similar \utsd{} solution
contains a self-similar \emph{sequence of supersonic patches} just behind the
leading triple point, each closed by its own small triple point and fan. Two
consequences are central: the canonical trajectory function $g(a_0)$ is a
numerically determined function of the single inner parameter and has no closed
form; and there is no single smooth leading-order fan for a closed-form first
correction to perturb.

% ===================================================================
\section{Formal Local First-Order Check: Operator and Adjoint}
\label{sec:matching}
% ===================================================================
The previous section rules out a closed-form perturbation of a single smooth
fan: the leading-order inner solution is the numerical Guderley patch cascade.
The calculation below is therefore used only as a local consistency check on
the linearisation and adjoint signs. It should not be read as a Fredholm theory
for the full mixed-type shock-reflection free-boundary problem.

On a smooth patch of a provisional background $u^{(0)},v^{(0)}$, writing
$u=u^{(0)}+\al\,w+\OO(\al^2)$, $v=v^{(0)}+\al\,z+\OO(\al^2)$ gives
\begin{equation}
  (u-1)u_\eta\ \longrightarrow\
  (u^{(0)}-1)\,\partial_\eta w + \bigl(\partial_\eta u^{(0)}\bigr)\,w,
  \label{eq:lin}
\end{equation}
so that the local correction operator has the form
\begin{equation}
  \mathcal L(w,z)=\Bigl((u^{(0)}-1)w_\eta+u^{(0)}_\eta w+z_\xi,\ \ w_\xi-z_\eta\Bigr).
  \label{eq:L}
\end{equation}
For the actual Guderley solution this operator must be supplemented by the
linearised Rankine--Hugoniot conditions on every shock, the perturbation of
sonic arcs and patch interfaces, and the variation of the self-similar domain.
Those data are numerical and are not replaced here by any local fan model.

\subsection{Formal adjoint and local null family}
The cancellation in the adjoint is useful enough to record explicitly.  For
smooth compactly supported perturbations, or equivalently modulo boundary and
shock-interface terms, consider
\begin{align}
  \langle (p,q),\mathcal L(w,z)\rangle
  &=\iint_\Omega p\bigl[(u^{(0)}-1)w_\eta+u^{(0)}_\eta w+z_\xi\bigr]
     +q(w_\xi-z_\eta)\,\dd\xi\dd\eta .
\end{align}
Integrating by parts gives the interior contribution
\begin{align}
  \iint_\Omega
  w\bigl[-(u^{(0)}-1)p_\eta-q_\xi\bigr]
  +z\bigl[-p_\xi+q_\eta\bigr]  \,\dd\xi\dd\eta,
\end{align}
because the term $p u^{(0)}_\eta w$ cancels the derivative of
$(u^{(0)}-1)$ arising from $\partial_\eta[p(u^{(0)}-1)]$.  The discarded
boundary term is
\begin{equation}
  \int_{\partial\Omega}
  \left[p(u^{(0)}-1)w n_\eta+p z n_\xi+q w n_\xi-q z n_\eta\right]ds,
  \label{eq:adjoint-boundary}
\end{equation}
with additional jump contributions on internal shock or patch interfaces.  The
formal differential adjoint is therefore
\begin{equation}
  \mathcal L^{*}(p,q)=\Bigl(-(u^{(0)}-1)p_\eta-q_\xi,\ \ -p_\xi+q_\eta\Bigr).
  \label{eq:Lstar}
\end{equation}
On the local fan background $u^{(0)}-1=-1/s^2$, $s=\xi/\eta$, the homogeneous
adjoint admits the one-parameter null family
\begin{equation}
  \hat u=p=\eta^{\sigma}\,s^{(\sigma+1)/2},\qquad
  \hat v=q=\eta^{\sigma}\,s^{(\sigma-1)/2},\qquad q=p/s,
  \label{eq:null}
\end{equation}
with residual identically zero. The exponent is denoted by $\sigma$ to avoid
confusion with the Hunter--Tesdall control parameter $a$ in \eqref{eq:a}; it
would be fixed by matching and boundary conditions in any complete local
problem.

\subsection{Implicit definition of the first correction}
\label{sec:H-def}
The leading trajectory is determined by the canonical function $g(a)$.  The
next correction is a different object.  With
\begin{equation}
  \delta=\sqrt{2(\Mach^2-1)},\qquad a=\frac{\alpha}{\delta},
  \label{eq:delta-a}
\end{equation}
the physical trajectory angle has the formal expansion
\begin{equation}
  \chi_{\rm phys}(a,\delta)
  =\delta\,g(a)+\delta^2 H(a;\gamma)+o(\delta^2),
  \label{eq:H-expansion}
\end{equation}
where $H$ denotes the first correction in the Mach-number strength convention.
We write $H(a;\gamma)$, rather than $h(a)$, because the leading \utsd{} problem
is $\gamma$-free but the next Euler/UTSD matching terms may contain outer
thermodynamic normalisations.  A $\gamma$-dependence at this order would
therefore be a next-order matching effect, not a leading-order inner dependence.

For a smooth fixed patch, the formal adjoint calculation gives the local
orthogonality condition
\begin{equation}
  \mathcal S_{\rm loc}
  =\iint_{\Omega_{\rm loc}}(F\hat u+G\hat v)\,\dd\xi\,\dd\eta
  +\text{boundary/interface contributions}=0.
  \label{eq:formal-solv}
\end{equation}
For the full Guderley free-boundary problem this bookkeeping identity should be
read schematically as a solvability condition for the correction amplitude.  If
$U_1=(u_1,v_1)$ is the first correction and $\mathcal L_a$ is the full
linearisation about the numerical Guderley solution at parameter $a$, then
\begin{equation}
  \mathcal L_a U_1=F_a+H(a;\gamma)R_a.
  \label{eq:full-linear-H}
\end{equation}
Here $F_a$ denotes the known forcing from the next-order asymptotic expansion,
while $R_a$ is the forcing produced by an infinitesimal change in the trajectory
angle.  Let $\Psi_a$ be the corresponding adjoint null vector and let
$\mathcal B_a$ denote all boundary and interface contributions from shocks,
sonic arcs, the wall, and Guderley patch interfaces.  The solvability condition
is
\begin{equation}
  \left\langle \Psi_a,F_a+H(a;\gamma)R_a\right\rangle
  +\mathcal B_a\bigl(F_a+H(a;\gamma)R_a\bigr)=0,
  \label{eq:H-solvability}
\end{equation}
and hence the first correction is defined implicitly by
\begin{equation}
  \boxed{\
  H(a;\gamma)=
  -\frac{\left\langle \Psi_a,F_a\right\rangle+\mathcal B_a(F_a)}
  {\left\langle \Psi_a,R_a\right\rangle+\mathcal B_a(R_a)}.\ }
  \label{eq:H-ratio}
\end{equation}
\revone{The denominator in \eqref{eq:H-ratio} is the adjoint projection of an
infinitesimal displacement of the trajectory.  For the coefficient to be
well-defined at this order one requires}
\begin{equation}
  \left\langle \Psi_a,R_a\right\rangle+\mathcal B_a(R_a)\ne0 .
  \label{eq:H-denominator}
\end{equation}
\revone{If this quantity vanished, the trajectory would not be fixed by the first
compatibility condition and a higher-order solvability condition would be
needed.}
Equation \eqref{eq:H-ratio} is the mathematically meaningful way to report the
first correction at this stage.  It specifies exactly what must be computed,
without replacing the full linearised free-boundary problem by the local fan
model.  The canonical time-marching solver in \Cref{sec:pilot} cannot extract
$H$, because it solves only the leading-order \utsd{} problem and contains no
independent finite-$\delta$ parameter.  An empirical estimate would require
full finite-$\delta$ Euler simulations, or a next-order \utsd{} solver, through
\begin{equation}
  H(a;\gamma)
  \approx
  \frac{1}{\delta}
  \left[\frac{\chi_{\rm phys}(a,\delta)}{\delta}-g(a)\right],
  \qquad \delta\to0.
  \label{eq:H-finite-delta}
\end{equation}
No numerical value or curve for $H$ is claimed here.

% ===================================================================
\section{Explicit Results}
\label{sec:results}
% ===================================================================
\subsection{The \utsd{} self-similar shock polar}
For a straight shock with unit normal
$n=(\cos\phi,\sin\phi)$ and normal speed $s$, the Rankine--Hugoniot condition
for the conservation law $u_t+(u^2/2)_x+v_y=0$ is
\begin{equation}
  -s[u]+\left[\frac12u^2\right]\cos\phi+[v]\sin\phi=0.
  \label{eq:rh-utsd-1}
\end{equation}
The compatibility equation $u_y-v_x=0$ gives the distributional jump condition
\begin{equation}
  [u]\sin\phi-[v]\cos\phi=0,
  \qquad\text{hence}\qquad [v]=[u]\tan\phi.
  \label{eq:rh-utsd-2}
\end{equation}
Substituting \eqref{eq:rh-utsd-2} into \eqref{eq:rh-utsd-1} and using
$[u^2]/2=\tfrac12(u_a+u_b)[u]$ gives
\begin{equation}
  s=\half(u_a+u_b)\cos\phi+\frac{\sin^2\phi}{\cos\phi}.
\end{equation}
In self-similar coordinates the normal speed of a straight line through
$P=(\xi_P,\eta_P)$ is $s=\xi_P\cos\phi+\eta_P\sin\phi$.  Thus the UTSD
self-similar shock polar is
\begin{equation}
  [v]=[u]\tan\phi,\qquad
  \xi_P\cos\phi+\eta_P\sin\phi
  =\half(u_a+u_b)\cos\phi+\frac{\sin^2\phi}{\cos\phi}.
  \label{eq:polar}
\end{equation}

\subsection{Regular reflection, detachment, and the window}
Normalise the state ahead of the incident shock to $(u_0,v_0)=(0,0)$ and
the state behind it to $u_1=1$.  If $t_i=\tan\phi_i$, the first jump condition
in \eqref{eq:polar} gives $v_1=t_i$.  For the reflected shock, joining state
$1$ to a wall state $2$ with $v_2=0$, write $t_r=\tan\phi_r$.  Then
\begin{equation}
  v_2-v_1=(u_2-u_1)t_r
  \quad\Longrightarrow\quad
  -t_i=(u_2-1)t_r,
  \qquad
  u_2=1-\frac{t_i}{t_r}.
  \label{eq:u2-wall}
\end{equation}
Since regular reflection occurs at the wall, the two shocks meet at the same
point $P=(\xi_P,0)$.  Dividing the second relation in \eqref{eq:polar} by
$\cos\phi$ gives, for a shock through the wall point,
\begin{equation}
  \xi_P=\half(u_a+u_b)+\tan^2\phi.
  \label{eq:wall-polar}
\end{equation}
For the incident shock this yields $\xi_P=\half+t_i^2$.  For the reflected
shock it yields $\xi_P=\half(1+u_2)+t_r^2$.  Substituting \eqref{eq:u2-wall}
and equating the two expressions for $\xi_P$ gives
\begin{equation}
  \half+t_i^2=1-\frac{t_i}{2t_r}+t_r^2.
\end{equation}
Multiplication by $2t_r$ gives the regular-reflection cubic
\begin{equation}
  2t_r^3+(1-2t_i^2)\,t_r-t_i=0.
  \label{eq:cubic-unfactored}
\end{equation}
It factors exactly as
\begin{equation}
  2t_r^3+(1-2t_i^2)\,t_r-t_i=(t_r-t_i)\bigl(2t_r^2+2t_i t_r+1\bigr),
  \qquad t_i=\tan\phi_i.
  \label{eq:cubic}
\end{equation}

\begin{figure}[t]
  \centering
  \includegraphics[width=0.74\linewidth]{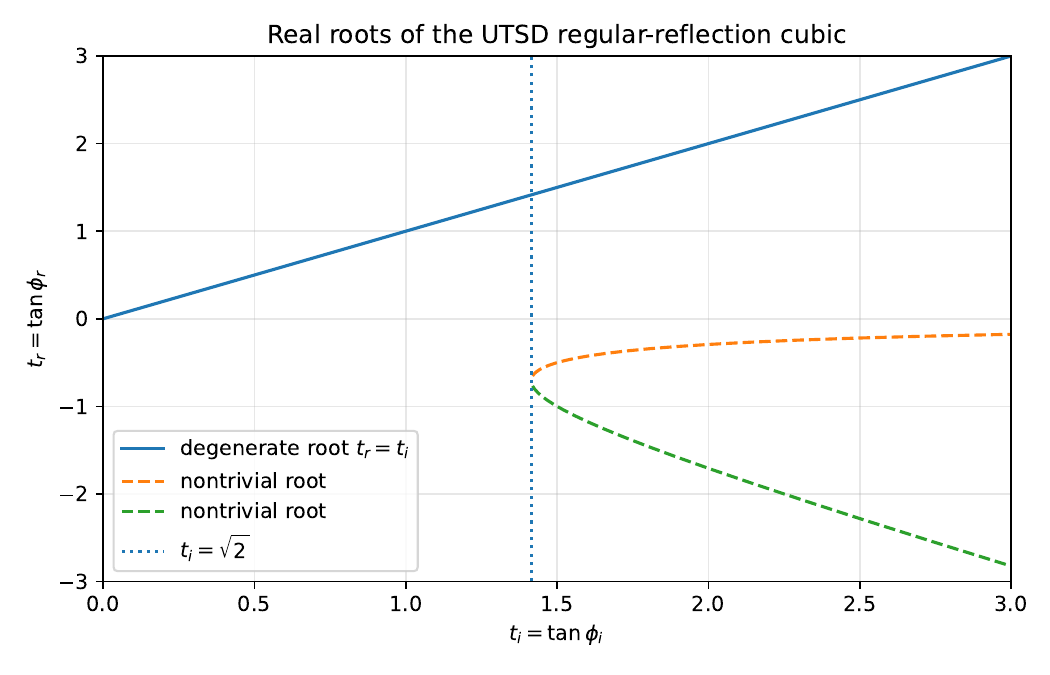}
  \caption{Real roots of the UTSD regular-reflection shock-polar cubic
  $2t_r^3+(1-2t_i^2)t_r-t_i=0$. The degenerate root $t_r=t_i$ exists for all $t_i$,
  whereas the nontrivial reflected-shock roots are real only for $t_i\ge\sqrt2$.
  This reproduces the detachment value $a_d=\sqrt2$ under the Hunter--Tesdall
  normalisation $t_i=a$.}
  \label{fig:shock-polar-roots}
\end{figure}

The nontrivial roots come from the quadratic factor,
\begin{equation}
  t_r=\frac{-t_i\pm\sqrt{t_i^2-2}}{2},
  \label{eq:tr-roots}
\end{equation}
and are real only for $t_i\ge\sqrt2$.  Below this value no real two-shock
solution exists; the root structure is shown in \Cref{fig:shock-polar-roots}.
This independently reproduces the detachment value
of the \utsd{} reflection problem,
\begin{equation}
  a_d=\sqrt2\approx1.414,
  \label{eq:ad}
\end{equation}
computed by \citet{HunterTesdall2004} (with nearby sonic value
$a_s\approx1.455$). Regular reflection occurs for $a>a_d$ and Guderley--Mach
reflection for $a<a_d$. Using $a=1/(2\sqrt\lam)$ from \eqref{eq:a}, the
admissible regimes in $\lam$ are
\begin{equation}
  \boxed{\text{RR:}\ \ a>\sqrt2\ \Longleftrightarrow\ \lam<\tfrac18,
  \qquad
  \text{Guderley--MR:}\ \ a<\sqrt2\ \Longleftrightarrow\ \lam>\tfrac18.\ }
  \label{eq:window}
\end{equation}
Alternative bounds such as $\lam>\half$ or $(\gamma+1)/2$ arise from the
normal-Mach-number parametrisation discussed in \Cref{sec:param}; the
consistent UTSD window is \eqref{eq:window}.
\revone{Physically, $\lambda$ measures shock strength relative to the square of the
glancing angle.  Increasing $\lambda$ therefore makes the shock stronger
relative to the geometry and decreases $a_0=1/(2\sqrt\lambda)$.  When
$a_0<\sqrt2$, the UTSD regular-reflection shock polar has no real nontrivial
reflected-shock branch satisfying the wall condition.  The two-shock regular
reflection then detaches and the leading-order inner solution must be the
Guderley--Mach configuration.  Thus $\lambda>1/8$ is not an additional physical
assumption; it is the Hunter--Tesdall detachment criterion expressed in the
chosen weak-shock/glancing scaling.}

\subsection{Leading-order trajectory angle}
By Proposition~2.1 and Section~2.4, the leading physical trajectory angle
is obtained from the canonical UTSD trajectory function $g(a_0)$ through
\begin{equation}
  \chi_{\rm phys}=2\sqrt{\lambda}\,\alpha\,g(a_0)+\OO(\alpha^3),
  \qquad a_0=\frac{1}{2\sqrt{\lambda}} .
  \label{eq:leading-angle}
\end{equation}
The function $g$ is obtained from the numerical solution of
\eqref{eq:ssUTSD}--\eqref{eq:IBVP}
\citep{HunterBrio2000,TesdallHunter2002,TesdallSandersKeyfitz2006}; it has no
elementary closed form. The next term is the implicit solvability coefficient
$H(a;\gamma)$ defined in \Cref{sec:H-def}; it is not evaluated numerically
here and no closed-form table is presented.

% ===================================================================
\section{Pilot Numerical Route to the Trajectory Function}
\label{sec:pilot}
% ===================================================================
The analysis above leaves one genuinely numerical object: the canonical
trajectory function $g(a_0)$ entering the physical angle through
\eqref{eq:leading-angle}. A complete computation of $g$ requires the full
self-similar mixed-type
Guderley solution.  As a reproducibility route, rather than as a converged
replacement for the Hunter--Tesdall calculation, we implemented a minimal
finite-volume time-marching discretisation of the canonical \utsd{} problem.
The system is advanced in the form
\begin{equation}
  u_t+\left(\frac12u^2\right)_x+v_y=0,\qquad v_x=u_y,
  \label{eq:pilot-evolve}
\end{equation}
with the oblique Riemann data \eqref{eq:IBVP}.  Thus the calculation is a
solver for the full leading-order canonical \utsd{} system in its
nonlocal time-dependent form: the transverse field $v$ is not prescribed or
neglected, but is recovered from the compatibility constraint $v_x=u_y$ at
each stage. It is therefore more than a one-dimensional Burgers calculation.
At the same time, it is not the full Hunter--Tesdall numerical construction: it
does not solve the steady self-similar mixed-type free-boundary problem with
shock fitting, sonic arcs, Guderley patch interfaces, and local adaptive
refinement. It should be read as a benchmarked time-marching route to the
leading-order canonical solution, not as a production-quality self-similar
Guderley solver.  At each stage $u_y$ is computed by centred differences and
the constraint $v_x=u_y$ is inverted by integrating from the right boundary,
using the far-field normalisation $v\to0$ as $x\to+\infty$:
\begin{equation}
  v(x,y,t)=-\int_x^{x_{\max}}u_y(s,y,t)\,\dd s,
  \qquad v(x_{\max},y,t)=0.
  \label{eq:v-reconstruction}
\end{equation}
Equivalently, the pilot calculation advances the scalar nonlocal equation
\begin{equation}
  u_t+\left(\frac12u^2\right)_x
  +\partial_y\left[-\int_x^{x_{\max}}u_y(s,y,t)\,\dd s\right]=0.
  \label{eq:nonlocal-utsd}
\end{equation}
The scalar Burgers flux in $x$ is treated by a Rusanov flux, while the nonlocal
transverse contribution is differenced conservatively in $y$.  The wall is
imposed through $v(x,0,t)=0$ and even reflection of $u$.  Crucially, $u$ is
not limited to $[0,1]$: the transverse coupling $v_y$ violates the scalar
maximum-principle intuition, so the flow behind the Mach stem is compressed to
$u>1$. Clipping at unity suppresses that compression and with it the Mach
reflection itself.

\begin{figure}[H]
  \centering
  \includegraphics[width=0.82\linewidth]{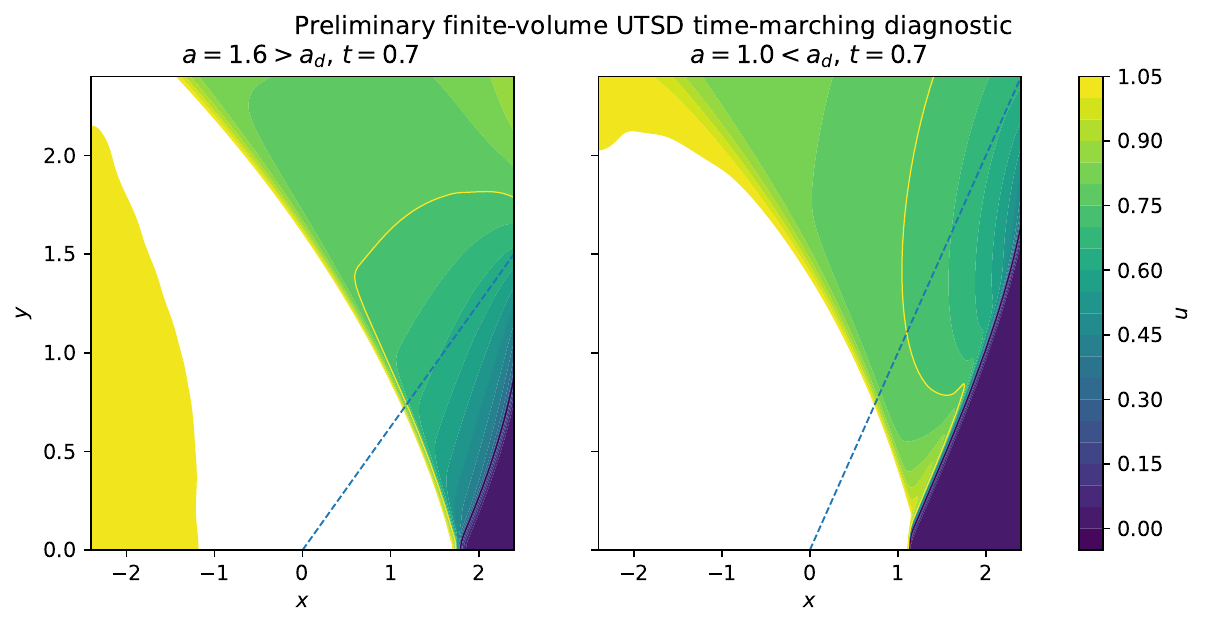}
  \caption{Pilot \utsd{} time-marching diagnostic.  The panels compare two
  canonical parameters, one above and one below the detachment value
  $a_d=\sqrt2$.  The diagnostic is not used as a converged Hunter--Tesdall
  solution; it documents the numerical route by which the trajectory function
  $g(a_0)$ can be recovered from late-time collapse in self-similar coordinates.}
  \label{fig:utsd-pilot}
\end{figure}

For a sequence of values of $a_0$, the late-time solution is plotted in
self-similar variables $(\xi,\eta)=(x/t,y/t)$, from which the shock contour
$\xi(\eta)$ can be extracted with sub-cell interpolation. As shown in
\S\ref{sec:pilot-validation}, this contour reproduces the published
Hunter--Tesdall triple-point location at $a_0=0.5$ once the $u>1$ compression
behind the Mach stem is retained. A naive global $|\nabla u|$ maximum, by
contrast, mislocates the triple point onto the near-wall foot of the Mach stem.
\revtwo{The purpose of \Cref{fig:utsd-pilot} is therefore limited but important: it
turns the leading-order canonical problem into a concrete, checkable
computation. It should not be read as a method that by itself determines a
reliable pointwise curve $g(a_0)$, nor as a validation of the full adaptive
self-similar Guderley free-boundary solution. Since the calculation solves only
the leading canonical \utsd{} equations, it cannot determine a next-order
coefficient $H(a;\gamma)$; that coefficient is defined only by the
linearised/free-boundary solvability problem described in \Cref{sec:H-def},
including perturbations of shocks, sonic boundaries, and Guderley patch
interfaces.}

\subsection{Self-similar collapse diagnostic}
\begin{figure}[H]
  \centering
  \includegraphics[width=0.78\linewidth]{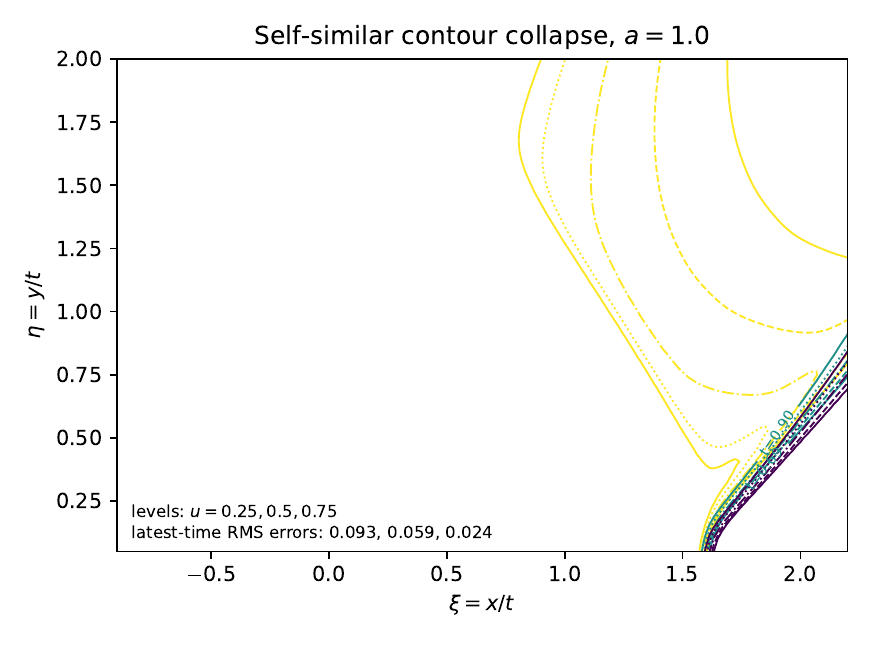}
  \caption{Late-time self-similar collapse diagnostic for the pilot UTSD
  solver at $a_0=1.0$.  Contours of $u$ at several times are plotted in
  $(\xi,\eta)=(x/t,y/t)$.  The quoted RMS errors are differences from the
  final snapshot after interpolation to a common similarity grid.}
  \label{fig:utsd-collapse}
\end{figure}

A necessary test for such a time-marching route is collapse in the similarity
variables $(\xi,\eta)=(x/t,y/t)$.  \Cref{fig:utsd-collapse} overlays the
contours $u=0.25,0.5,0.75$ for late-time snapshots at $a_0=1.0$.  Resampling
onto a common $(\xi,\eta)$ grid, the RMS difference relative to the final
snapshot decreases from $9.3\times10^{-2}$ to $5.9\times10^{-2}$ and then to
$2.4\times10^{-2}$ over the last three pre-final snapshots.  \revtwo{This is not a proof of convergence to the Hunter--Tesdall free-boundary
solution, but it is the basic numerical check that the evolutionary calculation
is approaching a self-similar leading-order structure rather than an arbitrary
transient.}

\subsection{Grid-stability diagnostic}
\begin{figure}[H]
  \centering
  \includegraphics[width=0.72\linewidth]{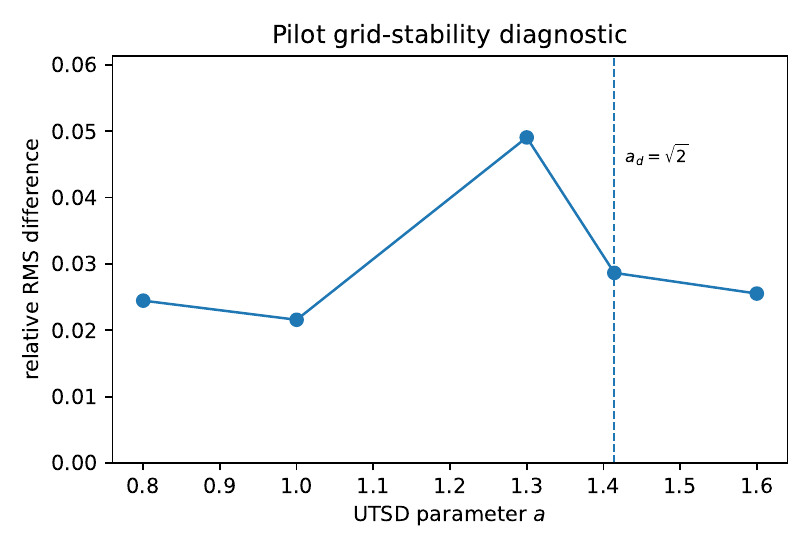}
  \caption{Pilot grid-stability diagnostic.  Each point is the relative RMS
  difference between $(140\times70)$ and $(210\times105)$ calculations,
  compared after interpolation to a common self-similar grid at $t=0.6$.
  The vertical dashed line marks $a_d=\sqrt2$.}
  \label{fig:utsd-grid}
\end{figure}

As a second reproducibility check, \Cref{fig:utsd-grid} compares low- and
medium-resolution runs after interpolation to the same similarity grid.  The
relative RMS differences are of order $2$--$5\%$ for the pilot parameters
shown.  The increase near $a_0\approx1.3$ is unsurprising for a first-order
shock-capturing method near the transition region, and should be reduced in a
production calculation by using a less diffusive reconstruction, local mesh
refinement near the reflection point, and a sharper extraction of the
triple-point geometry.  \revtwo{Thus the present numerics should be read as a demonstrated pilot workflow and
benchmark, not yet as a final table or curve for $g$.}

\subsection{Validation against Hunter--Tesdall and limitations of automatic angle extraction}
\label{sec:pilot-validation}
\citet{HunterTesdall2004} report the $a_0=0.5$ Mach-reflection triple point at
$(\xi,\eta)\approx(1.008,0.514)$, i.e. a trajectory angle
$\tan^{-1}(0.514/1.008)\approx27.0^\circ$. We use this point as a benchmark
for the time-marching solver.

\begin{figure}[H]
  \centering
  \includegraphics[width=0.78\linewidth]{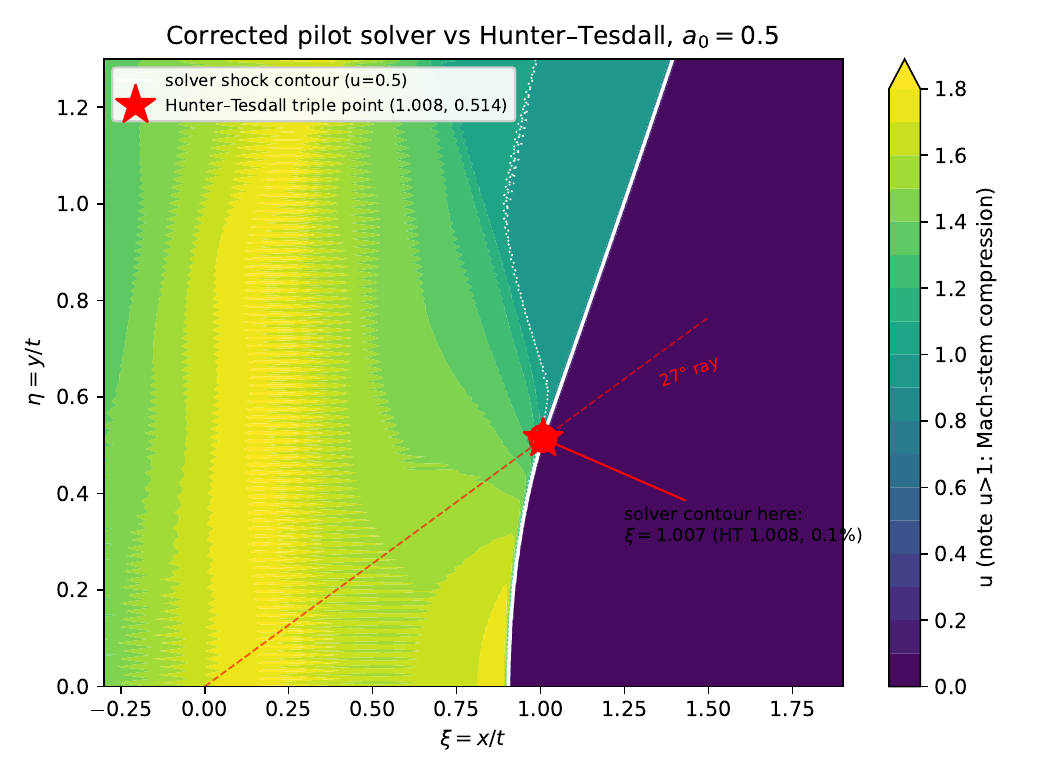}
  \caption{Pilot solver at $a_0=0.5$ compared with the Hunter--Tesdall triple
  point.  The $u=0.5$ shock contour passes through
  $(\xi,\eta)=(1.007,0.514)$, compared with the published
  $(1.008,0.514)$.  The dotted contour is $u=1$; the region $u>1$ behind the
  Mach stem is the transverse compression suppressed by any artificial
  $[0,1]$ clipping.}
  \label{fig:pilot-validation}
\end{figure}

The methodological point of \S\ref{sec:pilot} is essential here: the state must
not be limited to $[0,1]$. If the $u>1$ compression behind the Mach stem is
removed, the shock pattern collapses toward a single incident-shock line and
no Mach reflection is recovered. With $u>1$ admitted, the solver's $u=0.5$
shock contour passes through the published triple point. \Cref{tab:convergence}
lists the contour abscissa $\xi$ at $\eta=0.514$ at three resolutions, all
within $0.1\%$ of the Hunter--Tesdall value $\xi=1.008$; \Cref{fig:pilot-validation}
shows the corresponding field.

\begin{table}[H]
\centering
\caption{\revtwo{Benchmark at $a_0=0.5$: the pilot solver's $u=0.5$ shock-contour
abscissa at $\eta=0.514$ versus the Hunter--Tesdall triple-point value
$\xi=1.008$. This is a leading-order contour-location check for the canonical
time-marching solver, not a validation of a full adaptive self-similar Guderley
free-boundary computation.}}
\label{tab:convergence}
\begin{tabular}{lrr}
\toprule
grid & $\xi$ at $\eta=0.514$ & error vs HT ($1.008$)\\
\midrule
$400\times240$ & $1.007$ & $-0.1\%$\\
$560\times336$ & $1.007$ & $-0.1\%$\\
$720\times432$ & $1.007$ & $-0.1\%$\\
\bottomrule
\end{tabular}
\end{table}

Extracting a single trajectory angle from this field is more delicate than
locating the benchmark contour.  The Mach stem is curved: its local slope rises
smoothly from the wall toward the incident shock, so at pilot resolution there
is no uniquely defined kink at which an automatic algorithm should place the
triple point. A straight-line fit to the near-wall stem undershoots the apex,
whereas a fit near the apex merges into the incident shock. The global-maximum
estimator
\begin{equation}
  g_N(a_0)=\tan^{-1}(\eta_T/\xi_T),\qquad
  (\xi_T,\eta_T)=\arg\max |\nabla u|,
  \label{eq:gN}
\end{equation}
in particular locks onto the near-wall foot of the Mach stem rather than the
triple point.  Accordingly, we report the pilot calculation as a single
benchmarked anchor at $a_0=0.5$ and do not present an automatically swept
$g_N(a_0)$ curve.  A production-quality curve $g(a_0)$ would require adaptive
refinement near the triple point and sub-cell shock fitting, locating the
intersection of fitted incident, reflected, and Mach shock curves.

% ===================================================================
\section{Discussion}
\label{sec:discuss}
% ===================================================================
\subsection{The single-parameter, $\gamma$-free structure}
\begin{figure}[H]
  \centering
  \includegraphics[width=0.74\linewidth]{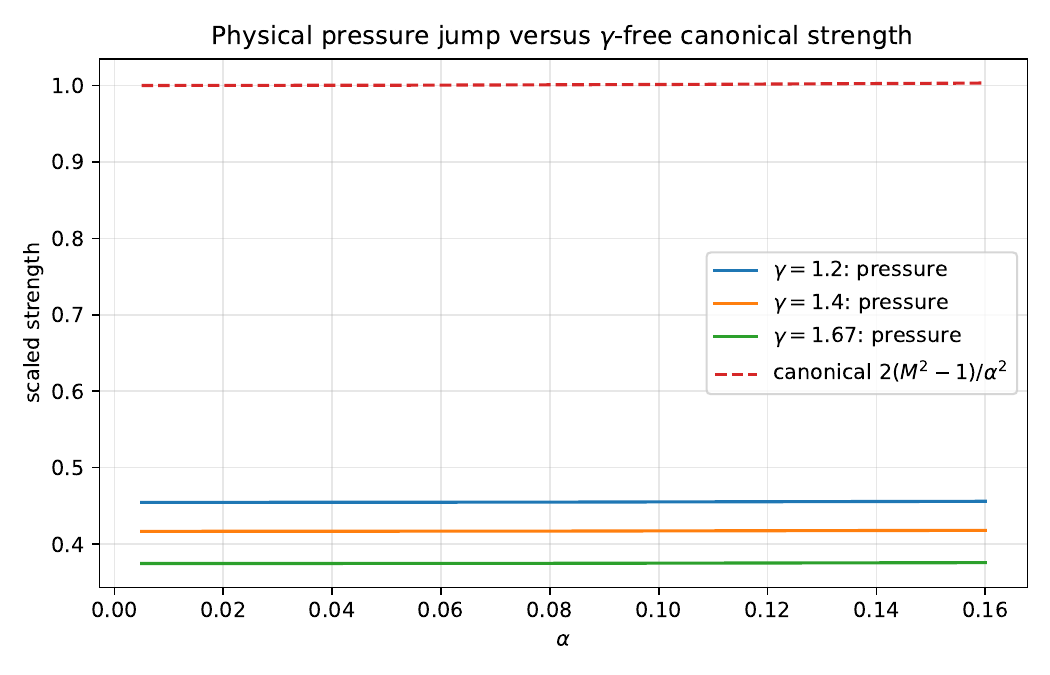}
  \caption{Diagnostic showing why $\gamma$ disappears from the canonical UTSD inner problem.
  The physical weak-shock pressure jump scaled by $\alpha^2$ depends on $\gamma$, whereas
  the canonical UTSD strength $2(\Mach^2-1)/\alpha^2$ is $\gamma$-free.}
  \label{fig:gamma-removal}
\end{figure}

The reduction to the single limiting $\gamma$-independent parameter $a_0=1/(2\sqrt\lam)$
(\Cref{prop:single}) is the central organising principle of the inner problem
and is consistent with the parametrisation of \citet{HunterTesdall2004}. The
two-parameter $(\gamma,\lam)$ space of the outer problem collapses, inside the
\utsd{} region, to a single $\gamma$-free axis. In the Mach-number strength
normalisation used here, the physical angle is recovered by the factor
$\sqrt{2(\Mach^2-1)}$ in \eqref{eq:chi-delta}; a $\gamma$-dependent factor appears
only if the strength is re-expressed using the pressure jump, as in
\eqref{eq:pressure-normalisation}. Any expression for a canonical inner quantity
as an independent rational function of $\gamma$ and $\lam$ is therefore
inconsistent with the structure of the equations. \Cref{fig:gamma-removal}
illustrates this separation at the level of the weak-shock scaling.

\subsection{The paradox regime and Guderley reflection}
The detachment value \eqref{eq:ad} marks the RR/MR transition. For $a<\sqrt2$
($\lam>\tfrac18$) the flow is a Guderley--Mach reflection with the
supersonic-patch cascade, and the leading-order triple-point data and any
corrections are properties of the numerical self-similar solution, consistent
with the numerical and experimental record
\citep{Skews2005,TesdallHunter2002,VasilevKraiko1999}. In the narrow band
$a_d<a<a_s$ the state behind the reflected shock is subsonic and the precise
transition mechanism is delicate \citep{HunterTesdall2004}.

\subsection{Extensions}
Real-gas thermodynamics affects the outer relations and the conversion between
pressure jump and Mach-number strength, but the inner problem is unchanged in
canonical form and still depends only on $a$. Wall curvature enters the inner
problem as a forcing term in \eqref{eq:L}. It may shift the numerically
determined $g(a)$ and any first-order correction, but it does not produce a
closed-form modification.

% ===================================================================
\section{Conclusions}
\label{sec:conclude}
% ===================================================================
\revone{We have clarified the distinguished scaling that connects weak shock reflection
at nearly glancing incidence to the known Hunter--Brio/Hunter--Tesdall \utsd{}
inner problem.} A single convention is
fixed; the incident-shock strength is $\Mach^2-1=2\lam\al^2+\OO(\al^4)$, and the ordinary post-shock flow deflection is $\OO(\alpha^3)$ with expansion \eqref{eq:outer-defl},
and the inner \utsd{} reflection problem is controlled, in the limit, by the
single, $\gamma$-independent parameter $a_0=1/(2\sqrt\lam)$, in agreement
with \citet{HunterTesdall2004}. Every canonical inner quantity depends on
$(\gamma,\lam)$ only through $\lam$, and the leading physical trajectory angle is
\begin{equation*}
  \chi_{\rm phys}=2\sqrt{\lambda}\,\alpha\,g(a_0)+\OO(\alpha^3),
\end{equation*}
with $g$ a function of the inner parameter alone. No separate $\gamma$-dependent
prefactor appears in this Mach-number-based scaling; such a factor is only a
change of normalisation if pressure jump is used as the strength variable. The
RR/MR transition is the detachment value
$a_d=\sqrt2$, which our independent shock-polar analysis reproduces, bounding
the Guderley--Mach regime as $\lam>\tfrac18$. The genuine leading-order inner
field is the mixed-type self-similar \utsd{} solution with Guderley
supersonic-patch structure, whose trajectory angle is determined numerically
and has no elementary closed form. The pilot time-marching calculation in
\Cref{sec:pilot} gives a practical route toward computing this function in the
canonical variables, and we benchmark it at $a_0=0.5$ against the
Hunter--Tesdall triple point $(1.008,0.514)$. The solver advances the full
leading-order canonical \utsd{} system, including the nonlocal reconstruction
of $v$ from $v_x=u_y$; it is not merely a Burgers-only approximation. Once the
transverse compression $u>1$ behind the Mach stem is admitted, the computed
$u=0.5$ contour passes through $(\xi,\eta)=(1.007,0.514)$, giving a $0.1\%$
match in the contour abscissa over the tested grids. Extracting a single
trajectory angle is more ambiguous on a coarse uniform grid because the curved
Mach stem merges smoothly into the incident shock. We therefore report the
pilot only as a single benchmarked anchor and do not present an automatically
swept $g_N(a_0)$ curve; a trustworthy point-by-point $g(a_0)$ requires adaptive
refinement and sub-cell shock fitting. The time-marching solver is also a
leading-order solver only, so it does not determine the first correction
coefficient $H(a;\gamma)$.  Instead \Cref{sec:H-def} defines $H$ implicitly
through the adjoint solvability condition for the full linearised Guderley
free-boundary problem.  Evaluating it would require the numerical Guderley
solution together with its linearised shock/interface conditions, or full
finite-$\delta$ Euler simulations. Consequently no numerical first-correction
curve, closed-form correction coefficient, critical parameter, or trajectory
table is claimed here.

% ===================================================================
\appendix
% ===================================================================
\newpage
\section{Supplementary local parabolic-fan calculation}
\label{app:parabolic-fan}
This appendix records an auxiliary local calculation from the parabolic fan
ansatz.  It is not used as a replacement for the genuine Guderley solution; its
purpose is to show why a local elementary fan does not close the higher-order
trajectory problem.

In local parabolic coordinates the UTSD system may be written as
\begin{equation}
  u_\tau+u u_\xi=v_\eta,\qquad v_\xi=u_\eta .
  \label{eq:local-utsd}
\end{equation}
Let
\begin{equation}
  s=\frac{\xi}{\eta^{1/2}},\qquad
  u_p(s)=1+\frac12 s,
  \label{eq:parabolic-fan}
\end{equation}
and define
\begin{equation}
  v_p(\xi,\eta)=\eta^{1/2}-\frac18 s^2\eta^{-1/2}
  =\eta^{1/2}-\frac{\xi^2}{8}\eta^{-3/2} .
  \label{eq:parabolic-v}
\end{equation}
Then
\begin{equation}
  (v_p)_\xi=(u_p)_\eta=-\frac14\xi\eta^{-3/2},
\end{equation}
so the compatibility equation in \eqref{eq:local-utsd} is satisfied exactly.
The dynamic equation is not satisfied by a static self-similar field alone:
\begin{equation}
  u_p (u_p)_\xi-(v_p)_\eta
  =\frac{s}{4}\eta^{-1/2}-\frac{3s^2}{16}\eta^{-3/2} .
  \label{eq:parabolic-residual}
\end{equation}
Thus \eqref{eq:parabolic-fan}--\eqref{eq:parabolic-v} should be interpreted as
a local matching and compatibility calculation.  The missing dynamic balance
must be supplied by the slow-time/geometric forcing or by the full self-similar
Guderley structure.

Linearising \eqref{eq:local-utsd} about this parabolic profile gives the model
operator
\begin{equation}
  \mathcal L_p\begin{pmatrix}w\\z\end{pmatrix}
  =\begin{pmatrix}
  u_p\,w_\xi+(u_p)_\xi w-z_\eta\\[2pt]
  -w_\eta+z_\xi
  \end{pmatrix},
  \qquad (u_p)_\xi=\frac12\eta^{-1/2}.
  \label{eq:parabolic-operator}
\end{equation}
With respect to the flat $L^2$ pairing, integration by parts gives the formal
adjoint
\begin{equation}
  \mathcal L_p^*\begin{pmatrix}\hat u\\ \hat v\end{pmatrix}
  =\begin{pmatrix}
  -u_p\,\partial_\xi\hat u+\partial_\eta\hat v\\[2pt]
  \partial_\eta\hat u-\partial_\xi\hat v
  \end{pmatrix}.
  \label{eq:parabolic-adjoint}
\end{equation}
For the weighted space
\begin{equation}
  \mathcal H_p=\left\{(w,z):\int_\Omega (w^2+z^2)\eta^{-1/2}\,d\xi d\eta<\infty\right\},
  \qquad \Omega=\{s_1\le s\le0,\ \eta>0\},
  \label{eq:parabolic-H}
\end{equation}
the weighted adjoint is $\mathcal L_{p,\mathcal H}^*=\rho^{-1}\mathcal L_p^*(\rho\cdot)$ with
$\rho=\eta^{-1/2}$, i.e.
\begin{equation}
  \mathcal L_{p,\mathcal H}^*\begin{pmatrix}\hat u\\ \hat v\end{pmatrix}
  =\begin{pmatrix}
  -u_p\,\partial_\xi\hat u+\partial_\eta\hat v-\frac{1}{2\eta}\hat v\\[4pt]
  \partial_\eta\hat u-\frac{1}{2\eta}\hat u-\partial_\xi\hat v
  \end{pmatrix}.
  \label{eq:parabolic-weighted-adjoint}
\end{equation}
Consequently $\ker \mathcal L_{p,\mathcal H}^*=\eta^{1/2}\ker\mathcal L_p^*$.

Within the power-law class $\hat u=\eta^m p(s)$,
$\hat v=\eta^n q(s)$, the flat adjoint kernel is
\begin{equation}
  \ker_{\rm ss}\mathcal L_p^*=\textrm{span}\{(1,0),\ (0,1),\ (\eta,\xi)\}.
  \label{eq:parabolic-kernel}
\end{equation}
Indeed substitution into \eqref{eq:parabolic-adjoint} gives
\begin{align}
  \eta^{n-1}\left(nq-\frac{s}{2}q'\right)
  -\eta^{m-1/2}\left(1+\frac{s}{2}\right)p'&=0,\label{eq:parabolic-adjA}\\
  \eta^{m-1}\left(mp-\frac{s}{2}p'\right)-\eta^{n-1/2}q'&=0.\label{eq:parabolic-adjB}
\end{align}
The two possible balance conditions, $m=n-1/2$ and $m=n+1/2$, cannot hold
simultaneously.  Separating the balanced and unbalanced cases leaves only the
two constant modes and the mode $(\eta,\xi)$, each of which is checked directly
in \eqref{eq:parabolic-adjoint}.

It follows that the corresponding self-similar weighted-adjoint modes are
\begin{equation}
  (\eta^{1/2},0),\qquad (0,\eta^{1/2}),
  \qquad (\eta^{3/2},\eta^{1/2}\xi),
  \label{eq:parabolic-weighted-modes}
\end{equation}
all of which are inadmissible in \eqref{eq:parabolic-H}: the weighted norm
integrands grow like $\eta^{1/2}$ or faster as $\eta\to\infty$.  Thus the
parabolic fan has no admissible self-similar adjoint annihilator in this
weighted space.  Since the outer forcing approaches an $O(1)$ state at the fan
edge, its pairing with these weighted modes is non-decaying over the
semi-infinite fan.  This is the local obstruction: a finite scalar solvability
condition, and hence an elementary closed form for the first correction, cannot
be obtained from this parabolic ansatz alone.

For reference, introduce the pressure-jump amplitude
\begin{equation}
  \Sigma=\frac{2\lambda}{\gamma+1},\qquad s_1=2(\Sigma-1).
\end{equation}
The elementary fan integrals that arise in such attempted reductions include
\begin{align}
\int_{s_1}^{0}\frac{s^2}{(1+s/2)^2}\,ds
  &=8\left(\frac1\Sigma-\Sigma+2\ln\Sigma\right),\label{eq:int1}\\
\int_{s_1}^{0}\frac{s^2}{1+s/2}\,ds
  &=8\left(-\frac32+2\Sigma-\frac{\Sigma^2}{2}-\ln\Sigma\right),\label{eq:int2}\\
\int_{s_1}^{0}\frac{s}{1+s/2}\,ds
  &=4(1-\Sigma+\ln\Sigma).\label{eq:int3}
\end{align}
Their logarithmic dependence is another sign that any genuine higher-order
coefficient, if computed from the correct Guderley free-boundary problem, should
not be expected to reduce to a simple rational function of $(\gamma,\lambda)$.
We emphasise that this non-closure is a statement about algebraic parabolic self-similar adjoint modes, not about the non-existence of an adjoint formulation altogether. A possible route beyond the present calculation is to introduce $(r=\eta^{1/2})$ and seek exponentially localised adjoint modes of the form
\begin{equation}
(\hat u,\hat v)=e^{-\mu r}r^\kappa\sum_{n\ge0}r^{-n}(P_n(s),Q_n(s)),\qquad \Re\mu>0.
\end{equation}
At leading order this produces a one-dimensional spectral problem on the fan interval,
\begin{equation}
(aP_0')'=\frac{\mu^2}{4}P_0,\qquad a=1+\frac{s}{2},
\end{equation}
with Bessel-type solutions. Such modes are ($H$)-admissible after the weighted-adjoint conjugation and therefore avoid the divergence of the algebraic modes. However, the parameter $(\mu)$ and the corresponding solvability condition are fixed only after imposing the full adjoint fan-edge, sonic-line, and inner-inner matching conditions. We therefore regard this as a route to a future higher-order theory rather than as part of the present closed-form leading-order result.

\section{Verification}
\label{app:verify}
The following were verified with a computer-algebra system (SymPy): the
strength, pressure-jump, and outer-deflection expansions \eqref{eq:pjump}--\eqref{eq:outer-defl} and the control
limiting parameter $a_0=1/(2\sqrt\lam)$ from \eqref{eq:a}; the similarity-exponent balance
and the parabolic compatibility field \eqref{eq:parabolic-fan}--\eqref{eq:parabolic-v}; the sonic line; the local linearisation \eqref{eq:lin},
formal adjoint \eqref{eq:Lstar}, and null family \eqref{eq:null} (residual
identically zero); and the polar and cubic \eqref{eq:polar}--\eqref{eq:cubic},
whose real-root threshold reproduces $t_i=\sqrt2=a_d$. The detachment value
$a_d=\sqrt2$ and the sonic value $a_s\approx1.455$ are taken from the
numerical/self-similar literature \citep{HunterTesdall2004,TesdallHunter2002}.
The pilot solver of \Cref{sec:pilot} is benchmarked against the Hunter--Tesdall
triple point at $a_0=0.5$ (\Cref{tab:convergence}); the match requires admitting
$u>1$ behind the Mach stem. The calculation advances the full leading-order
canonical \utsd{} time-dependent system with $v$ reconstructed from $v_x=u_y$,
but it is not a direct adaptive self-similar free-boundary calculation. The
global-maximum estimator \eqref{eq:gN} mislocates the triple point, so no
automatically swept $g_N(a_0)$ curve is reported. No value of $H(a;\gamma)$ is claimed here, since that coefficient belongs to
the next-order linearised/free-boundary problem.

% ===================================================================

\end{document}